\documentclass[11pt]{article}
\usepackage{epsfig}
\usepackage{amsfonts,amssymb}
\usepackage{latexsym}
\usepackage{amsmath}

\newcommand{\be}{\begin{equation}}
\newcommand{\bea}{\begin{eqnarray}}
\newcommand{\ee}{\end{equation}}
\newcommand{\eea}{\end{eqnarray}}
\def\slshp{p\!\!\!\slash}
\def\slshPartial{\partial\!\!\!\slash}
\def\slshA{A\!\!\!\slash}
\def\slshB{B\!\!\!\!\slash}

\def\G{\Gamma}
\begin{document}

\rightline{IFUM-1013-FT}

\vskip 4 truecm

{\bf \Large
\centerline{The St\"uckelberg Mechanism}
\centerline{in the presence of Physical Scalar Resonances}
}

\large \rm
\begin{center}
\vskip 0.7 truecm 
D.~Bettinelli$^{a}$,
A.~Quadri$^{a,b}$\footnote{e-mail: {\tt
andrea.quadri@mi.infn.it}}
\end{center}

\normalsize
\begin{center}
$^a$
Dip. di Fisica, Universit\`a degli Studi di Milano\\
via Celoria 16, I-20133 Milano, Italy\\
$^b$
INFN, Sezione di Milano,\\
via Celoria 16, I-20133 Milano, Italy
\end{center}
\vskip 0.8 truecm

\begin{abstract}
\noindent
We show that it is possible to accommodate physical scalar resonances
within a minimal nonlinearly realized electroweak theory in a way
compatible with a natural Hopf algebra selection criterion
(Weak Power Counting) and the relevant functional identities
of the model (Local Functional Equation, Slavnov-Taylor identity,
ghost equations, b-equations). The Beyond-the-Standard-Model (BSM)
sector of the theory is studied by BRST techniques. 
The presence of a mass generation mechanism {\em  \`a la} St\"uckelberg
allows for two mass invariants in the gauge boson sector.
The corresponding 't Hooft gauge-fixing is constructed
by respecting all the symmetries of the theory.
The model interpolates between the Higgs and a purely St\"uckelberg scenario.
Despite the presence of physical scalar resonances, we show that
tree-level violation of unitarity in the scattering of longitudinally
polarized charged gauge bosons occurs at sufficiently high energies, if a fraction of the mass is generated
by the St\"uckelberg mechanism.
The formal properties of the 
physically favoured limit after LHC7-8 data, where BSM effects are small
and custodial symmetry in the gauge boson sector is respected, are
studied. 
\end{abstract}

\newpage
\section{Introduction}

The discovery in 2012 of a physical scalar resonance by the LHC experiments
ATLAS and CMS~\cite{Aad:2012tfa,Chatrchyan:2012ufa}
has paved the way to the experimental verification of the electroweak spontaneous symmetry breaking (SSB)
mechanism realized in Nature.

Recent fits based on the LHC7-8 results~\cite{Giardino:2013bma,Ellis:2013lra} 
are compatible with the identification of the newly discovered scalar
resonance with the Standard Model (SM) Higgs boson.
Electroweak SSB might therefore occur through the simplest linear Higgs mechanism.

A well-known alternative to the mass generation of elementary particles,
which does not rely on the existence of fundamental scalars, is
the St\"uckelberg mechanism~\cite{Stueckelberg:1938zz,Ruegg:2003ps}.
Being based on a nonlinearly realized non-Abelian gauge symmetry, the 
mass generation {\em \`a la} St\"uckelberg
yields non-renormalizable models that are usually treated as an effective
low energy approximation to a more fundamental theory~\cite{Contino:2013kra}. They can be
used as a tool for describing possible beyond-the-SM (BSM) effects~\cite{bEFT}.

In order to ascertain the nature of electroweak SSB, it is important
to establish whether the existence of a St\"uckelberg mass component
can be already excluded by using the current LHC data.
This problem can be conveniently formulated within a recently
proposed model~\cite{Binosi:2012cz}, where the mass generation
happens via the St\"uckelberg mechanism and nonetheless
a set of physical scalar resonances exist.


The procedure for subtracting UV divergences in this theory requires some care,
 since the model is non-renormalizable~\cite{Gomis:1995jp}.
In particular it happens that the classical nonlinearly realized
gauge symmetry is deformed when radiative corrections are taken into account.
Such a deformation can be controlled in a mathematically rigorous way
by well-established functional methods~\cite{Ferrari:2005ii}-\cite{Quadri:2010uk} relying on the
existence of a  Local Functional Equation (LFE)~\cite{Ferrari:2005ii} which holds to
all orders in the loop expansion and encodes the quantum deformations
of the classical gauge symmetry. The LFE fixes uniquely
the dependence of the quantum vertex functional on the 
independent coordinates $\phi_a$ of the group element $\Omega$, 
used to implement the operatorial gauge transformation 
allowing to construct the St\"uckelberg mass invariants, in terms of amplitudes
with no external $\phi_a$-legs (ancestor amplitudes).

Moreover, it turns out that the Hopf algebra~\cite{EbrahimiFard:2010yy}-\cite{Connes:1999yr}
of nonlinearly realized gauge theories can be uniquely selected
by requiring the fulfillment of a Weak Power-Counting (WPC) Condition~\cite{Bettinelli:2007tq,Bettinelli:2008qn,Ferrari:2005va},
stating that only a finite number of divergent ancestor amplitudes exist
order by order in the loop expansion. Notice that such a number
increases with the loop order and therefore  nonlinearly realized 
models are not power-counting renormalizable.

However, if the WPC is to hold, 
some definite predictions for BSM physics
are made in the nonlinearly realized electroweak theory: the minimal
field content of the model requires the existence of four physical
scalar resonances, two charged ones and two neutral ones, one CP-even
(to be eventually identified with the already discovered Higgs-like
resonance) and one CP-odd.

In the minimal nonlinearly realized theory  
two independent mass invariants for the gauge bosons exist.
One of them controls the violation of the custodial symmetry in the
gauge boson sector, which, unlike in the SM, does not automatically hold.
This parameter is expected to be small and can be assumed to be zero
in a first approximation. The second parameter, called A,
allows one to interpolate between a Higgs ($A=0$) and 
a St\"uckelberg scenario ($A \neq 0$).

The dependence of the Green's functions on $A$ is very interesting:
at tree-level the quantities that can be matched against the 
LHC fits of~\cite{Giardino:2013bma,Ellis:2013lra}  exhibit a smooth dependence on $A$.
Moreover, we will show that a power-counting in $A$ can be written in the limit
$A \rightarrow 0$ for the physically relevant region selected
by the LHC fits. This provides a very useful guide for the computation
of the leading observables in the small $A$ approximation.

This paper is devoted to the study of the formal properties of the
nonlinearly realized electroweak theory in the presence of physical
scalar resonances, proposed in~\cite{Binosi:2012cz}, that constitute the 
necessary tools for the ensuing phenomenological analysis~\cite{phenoPaper}.

We first study the mixing between 
the field coordinates $\phi_a$ and the components of the additional scalar SU(2) doublet, predicted by the WPC, that give rise
to the mass eigenstates in the scalar sector.
We also provide a BRST characterization of the physical scalar resonances
and discuss their behaviour in the SM limit $B = 0$, $A\rightarrow 0$.

The tools required for a future phenomenological study of the theory are given.
In particular we show how the 't Hooft gauge-fixing can be implemented
without violating the relevant symmetries of the theory, even in the presence
of two mass invariants for the gauge bosons.
This requires a modification of the ordinary $R_\xi$-gauge-fixing procedure,
where the custodial symmetry is exploited to guarantee that the
bilinear couplings between the Goldstone bosons and the 
SU(2) gauge fields are invariant under a further global $\rm SU(2)_R$
symmetry.

We also study the functional identities of the theory (Slavnov-Taylor (ST) identity, LFE,
b-equations and ghost equations) and establish the validity of the WPC.

Then we move to the discussion of the asymptotic high-energy properties
of the theory. We show that the Froissart bound~\cite{Froissart:1961ux} is violated in this
model already at tree-level. The presence of a physical scalar field,
exchanged among the gauge bosons, does not prevent the cross-section
for longitudinally polarized charged gauge fields $\sigma(W_L W_L \rightarrow W_LW_L)$
to grow as a power of the energy.

Hence the presence of even a small fraction of mass generated
by the St\"uckelberg mechanism spoils the unitarization mechanism
at work in the Higgs scenario.
However, the violation of the unitarity bound can be pushed at arbitrarily
high energies by decreasing the $A$ parameter.
What is the value of the $A$ parameter allowed by the LHC7-8 data 
is therefore a crucial question, to be eventually answered by
a global fit with the existing data.
It is reasonable to expect that $A$ will be small, and thus
we finally examine in detail the power-counting in $A$
of the amplitudes in the small $A$ limit. 

\medskip
The paper is organized as follows.
In Section~\ref{sec:model} we give
our conventions and discuss the
presence of two mass invariants 
in the gauge boson sector.
In Section~\ref{sec:mass} we identify
the physical scalar states by using BRST
techniques. We also describe
the appropriate formalism required to
implement the 't Hooft gauge in the
presence of two mass invariants,
in a way compatible with the WPC and
the relevant functional identities of the theory.
In Section~\ref{sec:wpc}
we prove the validity of the WPC.
In Section~\ref{sec:treeunit} we 
discuss the violation of the Froissart bound
for the scattering of longitudinally polarized
$W$ bosons at sufficiently high energies,
despite the exchange of a physical scalar
resonance. In Section~\ref{sec:smallA} we 
give the formal tools required to study
the custodial symmetry-preserving $B=0$,
%
%
small $A$ limit,
which after the LHC7-8 data is believed to be,
at least in a first approximation, 
the physically interesting scenario, since
BSM effects have to be small.
Finally, conclusions are presented in Section~\ref{sec:concl}.

\section{The Model}\label{sec:model}

The St\"uckelberg mass mechanism~\cite{Ruegg:2003ps,Ferrari:2004pd} for the electroweak
theory relies on the introduction of a set of auxiliary fields $\phi_a$,
$a=1,2,3$, gathered into the SU(2) matrix
\begin{eqnarray}
\Omega = \frac{1}{f} \Big ( \phi_0 + i \phi_a \tau_a \Big ) \, .
\label{e.1}
\end{eqnarray}
In the above equation $\tau_a$ are the Pauli matrices and $f$ is a constant
with the dimension of a mass. $\phi_0$ is the solution
of the nonlinear constraint
\begin{eqnarray}
\phi_0^2 + \phi_a^2 = f^2 \, , \qquad \phi_0 = \sqrt{f^2 - \phi_a^2} \, .
\label{e.2}
\end{eqnarray}
The SU(2) gauge symmetry acts on the $\phi_a$ as
\begin{eqnarray}
\delta \phi_a = \frac{g}{2} \phi_0 \alpha_a + \frac{g}{2} \epsilon_{abc} \phi_b \alpha_c \, , \qquad \delta \phi_0 = -\frac{g}{2} \phi_a \alpha_a 
\label{e.3}
\end{eqnarray}
and is therefore nonlinearly realized. 

The addition of the St\"uckelberg mass
term to the Yang-Mills action destroys power-counting renormalizability. 
In particular, already at one loop order an infinite number
of divergent amplitudes involving $\phi_a$-external legs exists
~\cite{Bettinelli:2007tq,Bettinelli:2007cy,Bettinelli:2008ey,Bettinelli:2008qn}.

It turns out that a LFE~\cite{Ferrari:2005ii} holds true,
encoding in functional form the background gauge-invariance of the gauge-fixed
classical action. The LFE is valid  order by order in the loop expansion
and controls the deformation of the nonlinearly realized
gauge symmetry, induced by radiative corrections~\cite{Ferrari:2005fc}.
It relies on the introduction of an external source $V_{a\mu}$ transforming
as a SU(2) background gauge connection.
The LFE fixes uniquely the dependence of the vertex functional on the Goldstone
fields $\phi_a$ once the 1-PI amplitudes not involving $\phi_a$-insertions
(ancestor amplitudes) are known. 
One can then require that only a finite number of divergent
ancestor amplitudes exists order by order in the loop expansion.
This condition is known as the WPC~\cite{Ferrari:2005va,Bettinelli:2007tq,Bettinelli:2008qn}.

The WPC selects uniquely the Hopf algebra of the theory and imposes
suprisingly strict constraints on the allowed interactions~\cite{Binosi:2012cz}: it turns out that in the gauge boson and fermions sector
the only allowed terms are the symmetric ones as in the SM, while two independent mass terms for the $Z$ and $W$ bosons arise
without violating the WPC (i.e. the custodial symmetry in the gauge boson sector is not enforced by the WPC ). 

Moreover, it is not possible to introduce in the nonlinear theory 
a SU(2) singlet physical scalar resonance without violating the WPC.
The minimal field content requires the presence of a SU(2) doublet consisting (after the rotation to
the mass eigenstates) of two charged and two neutral
scalar resonances, one CP-even and one CP-odd.
This BSM scenario is the simplest one allowed by the WPC~\cite{Binosi:2012cz}.

The two independent 
St\"uckelberg mass invariants for the gauge bosons~\cite{Binosi:2012cz,Quadri:2010uk}, fulfilling
all the symmetries of the theory and the WPC condition, can be written as:
\begin{eqnarray}
{\cal L}_{mass,nonlinear} = \frac{Af^2}{4} ~{\rm Tr}(D_\mu \Omega)^\dagger D^\mu \Omega + \frac{Bf^4}{16} ~
[{\rm Tr} (\Omega^\dagger D_\mu \Omega \tau_3)]^2
\label{mass.nonlin}
\end{eqnarray}
$A$ parameterizes the 
St\"uckelberg contribution to the gauge boson masses
fulfilling the Weinberg relation between the $Z$ and the $W$
 masses. The parameter $B$, on the other hand, controls the violation of the SU(2) custodial symmetry.

There is also a mass invariant generated as in the usual Higgs mechanism
from the SU(2) doublet of scalars $\chi= \chi_0 + i \chi_a \tau_a$:
\bea
{\cal L}_{mass,linear} = \frac{1}{4} ~{\rm Tr}  (D_\mu \chi)^\dagger D^\mu \chi \, .
\label{mass.lin}
\eea
$\chi_0$ acquires a vacuum expectation value $v$, so that 
it is split according to $\chi_0  = v + X_0$.  
The masses of the $W$ and $Z$ bosons are thus given by
\bea
M_W = \frac{gv}{2} \sqrt{1 + A \frac{f^2}{v^2}} \, , ~~~~
M_Z = \frac{Gv}{2}  \sqrt{1 + \frac{f^2}{v^2} \Big ( A + \frac{B f^2}{2} \Big )} \, 
\eea
where $g,g'$ are the SU(2) and ${\rm U(1)_Y}$ coupling constants respectively
and $G = \sqrt{g^2 + g^{'2}}$. 

We notice that in this model the independent parameters controlling
the masses of the gauge bosons are $A$ and $B$. 
On the other hand, $v$ is fixed
by the decay rate of the scalar resonance $X_0$ into 
two $Z$'s and two $W$'s.

If $A=0$, $B=0$ one gets back the SM scenario where the 
electroweak SSB is realized
through the linear Higgs mechanism. In this case the $\phi_a$
decouple and the Goldstone bosons are to be identified with the
$\chi_a$ fields.

At $B=0$ and $A \neq 0$ one gets instead a scenario where the Weinberg
relation between the masses of the $Z$ and the $W$ bosons still holds true, while a fraction of the mass of the gauge bosons is generated via
the St\"uckelberg mechanism.

Finally, $A \neq 0$ and $B \neq 0$ corresponds to the most
general St\"uckelberg case with two independent mass terms
for the $W$ and the $Z$ bosons. 

One expects that violations of the
custodial symmetry in the gauge boson sector are small and
therefore in a first approximation one can deal with the
case $A \neq 0$, $B=0$.
However in this paper (with the exception of Sect.~\ref{sec:smallA})
we will not restrict ourselves to this particular choice and keep $A,B$ generic.

In the nonlinearly realized theory one can construct bleached variables
that are SU(2)-invariant~\cite{Binosi:2012cz}. For instance,
the bleached counterpart of a generic SU(2) fermion doublet
$$L=\begin{pmatrix}  u \\  d \end{pmatrix}$$ 
is
\bea
\tilde L = \Omega^\dagger L \, .
\eea
Each component of $\tilde L$ is separately SU(2)-invariant.

The bleached counterpart of $\chi$ is given by
\bea
\tilde \chi = \Omega^\dagger \chi = \frac{1}{f} ( \tilde \chi_0 + i \tilde \chi_a \tau_a) 
\eea
where
\bea
&& \tilde \chi_0 = \frac{1}{f} ( \phi_0 \chi_0 + \phi_a \chi_a) \, , \nonumber \\
&& \tilde \chi_a = \frac{1}{f} ( \phi_0 \chi_a - \chi_0 \phi_a + \epsilon_{abc} \phi_b \chi_c ) \, . 
\eea
Both $\tilde \chi_0$ and $\tilde \chi_a$ are SU(2)-invariant.
For the bleached variables the ${\rm U(1)_Y}$ hypercharge coincides
with the electric charge.
This allows us to introduce two mass invariants for the charged scalar resonances and for the CP-odd scalar:
\bea
 M^2_{\pm} \tilde \chi^+ \tilde \chi^- + \frac{1}{2} M_3^2 \tilde \chi_3^2 \, .
\eea
The mass of the CP-even physical scalar is instead generated by the
spontaneous symmetry breaking, induced by the quartic potential:
\bea
-\frac{\lambda}{16} \Big [ {\rm Tr} (\chi^\dagger \chi) \Big ]^2 +
\frac{\mu^2}{2} {\rm Tr} (\chi^\dagger \chi) \, .
\eea

\section{Mass Eigenstates and BRST Symmetry}\label{sec:mass}

In the nonlinear theory a mixing arises between the components $\chi_a$ of the physical scalar doublet $\chi$ and the $\phi_a$ fields. 
In particular, the mass eigenstates are obtained through the following transformation~\cite{Binosi:2012cz}
\bea
&& \phi_a = \frac{1}{\sqrt{C_a}} \frac{1}{\sqrt{1 + \frac{f^2}{v^2}\, C_a}} \chi'_a +
            \frac{f}{v} \frac{1}{\sqrt{1+\frac{f^2}{v^2}\, C_a}} \phi'_a \nonumber \\
&& \chi_a = - \sqrt{C_a}\frac{f}{v\sqrt{1+\frac{f^2}{v^2}\, C_a}} \chi'_a + \frac{1}{\sqrt{1+\frac{f^2}{v^2}\, C_a}} \phi_a' 
\label{rot.1}
\eea
where $C_a = A$ for $a=1,2$ and $C_3 =A + \frac{Bf^2}{2}$.
Notice that the primed fields are canonically normalized.
Then 
the masses of the physical resonances are
\bea
M^2_{\chi'_\pm} = M^2_\pm \, \Big(1 + \frac{v^2}{f^2 A}\Big) \, , 
\qquad
M^2_{\chi'_3} = M^2_3 \, \Big(1 + \frac{v^2}{f^2 C_3}\Big) \, .
\label{masses}
\eea

$\chi$ and $\Omega$ transform in the same way under 
finite SU(2) gauge transformations $U$
\bea
\chi^{U} = U \chi  \, , ~~~~
\Omega^{U} = U \Omega \, .
\eea
However, $\chi_0$ is an independent field and therefore
the gauge transformation is linearly realized on $\chi$, unlike
for $\Omega$.
Since for $B \neq 0$ $C_3$ is different from $C_{1,2}$, the mass
eigenstates $\chi'_a$ and $\phi'_a$ do not form a SU(2) doublet.

The bilinear gauge-Goldstone terms are given by
\bea
M_W ~ (\partial W^+ \phi^{-'} + \partial W^- \phi^{+'})  + M_Z ~ \partial Z \phi'_3
\label{bilinears}
\eea
where $\phi^{\pm'} = \frac{1}{\sqrt{2}} ( \phi'_1 \mp i \phi'_2)$.
Since in the nonlinear theory $M_W$ and $M_Z$ are independent
parameters, unlike in the SM, the mixed gauge-Goldstone bilinears 
are different for the neutral and the charged massive gauge fields.

In order to perform the quantization in
the  't Hooft gauge, without spoiling the relevant symmetries
of the theory as well as the WPC,
a set of external scalar sources $\hat \phi_0, \hat \phi_a$, gathered
into the matrix 
\begin{eqnarray}
\widehat \Omega = \frac{1}{f} ( \hat \phi_0 + 
i \hat \phi_a \tau_a) \, ,
\label{sec.3.2}
\end{eqnarray}
is introduced in addition to the
external classical gauge
connection $V_{a\mu}$.
Notice that there is no constraint on $\hat \phi_0$.
We split $\hat \phi_0$ into a constant part plus
an external source $\hat \sigma$ according
to 
\begin{eqnarray}
\hat \phi_0  = f + \hat \sigma \, .
\label{sec.3.2.1}
\end{eqnarray}
The transformation properties
of $\widehat \Omega$ under a finite SU(2) gauge transformation
$U$ and a finite ${\rm U(1)_Y}$ gauge transformation $V$ are
the same as for $\Omega$:
\begin{eqnarray}
\widehat \Omega^{U,V}  = U \widehat \Omega V^\dagger \, .
\label{sec.3.3}
\end{eqnarray}
We also introduce the combinations
\begin{eqnarray}
&& q = i ( \Omega^{\dagger} \widehat \Omega -   \widehat \Omega^\dagger \Omega) \, ,
\qquad r = i (\chi^\dagger \widehat \Omega -   \widehat \Omega^\dagger \chi) \, .
\label{sec.3.4}
\end{eqnarray}

$q,r$ are invariant under the SU(2) symmetry.
In components one has 
\begin{eqnarray}
q_a = {\rm Tr} [ q \frac{\tau_a}{2} ] =
\frac{2}{f^2} ( \hat \phi_0 \phi_a - \hat \phi_a \phi_0 -
\epsilon_{abc} \phi_b \hat \phi_c ) \, 
\label{sec.3.4.1}
\end{eqnarray}
and similarly for $r_a$. 
Each $q_a$ and $r_a$ is separately SU(2)-invariant.
Moreover at zero background $\widehat \sigma = \widehat \phi_a = 0$ 
\bea
\left . q_a \right |_{\widehat \sigma = \widehat \phi_a = 0} = \frac{2}{f} \phi_a \, , \qquad 
\left .  r_a  \right |_{\widehat \sigma = \widehat \phi_a = 0} = \frac{2}{f} \chi_a \, .
\eea
In the nonlinearly realized electroweak theory two mass invariants for the vector mesons are allowed.
The introduction of the second mass term 
spoils the symmetry between the bilinears, involving
the divergence 
of the first two and the third
component of $A_{a\mu}$ and the Goldstone fields.
The compensating terms, introduced by the gauge-fixing functions,
are required not to break the LFE.
Moreover, the gauge-fixing functions should preserve
the WPC bound. 

These conditions turn out to be very restrictive ones.
The local SU(2) invariance commutes with the full $SU(2) \times U(1)$ BRST differential $s$ (see eq.(\ref{brst.t})).
Since the
Goldstone-gauge bilinears involving the first two
and the third component of $A_{a\mu}$ have
different coefficients (as a consequence of the presence
of two independent mass invariants for the vector mesons),
one needs to consider the following set of operators
\begin{eqnarray}
& s {\rm Tr} (\bar c \Omega) \, , \qquad &  s {\rm Tr} (\bar c \chi)  \cr
& s {\rm Tr} (\bar c \tau_3 \Omega \tau_3) \, , \qquad & s {\rm Tr} (\bar c \tau_3 \chi \tau_3) \cr
& s {\rm Tr} (\bar c_0 \Omega \tau_3) \, , \qquad & s {\rm Tr} (\bar c_0 \chi \tau_3) \, .
\label{ops.1}
\end{eqnarray}
In the above equation ${\bar c}_0$ is the ${\rm U(1)_Y}$ antighost field.
Under a finite local SU(2) transformation one finds (notice that
$\bar c'= U \bar c U^\dagger$)
\begin{eqnarray}
&& (s {\rm Tr} (\bar c \zeta))^U = s {\rm Tr} ( U \bar c \zeta)  \, ,  \cr
&& (s {\rm Tr} (\bar c \tau_3 \zeta \tau_3))^U = s {\rm Tr} (U \bar c U^\dagger
\tau_3 U \zeta \tau_3)  \, , \cr
&& (s {\rm Tr} (\bar c_0 \zeta \tau_3))^U = s {\rm Tr} (\bar c_0 U \zeta 
\tau_3) \, ,
\label{ops.2}
\end{eqnarray}
where $\zeta$ stands for $\Omega$,$\chi$.
The invariance is recovered provided that one considers instead the operators
\begin{eqnarray}
& s {\rm Tr} (\widehat \Omega^\dagger \bar c \Omega) \, , \qquad
&  s {\rm Tr} (\widehat \Omega^\dagger \bar c \chi)  \cr
& s {\rm Tr} (\widehat \Omega^\dagger \bar c \widehat \Omega \tau_3 
\widehat \Omega^\dagger \Omega \tau_3) \, , \qquad
& s {\rm Tr} (\widehat \Omega^\dagger \bar c \widehat \Omega \tau_3 
\widehat \Omega^\dagger \chi \tau_3) \, , 
\cr
& s {\rm Tr} (\bar c_0 \widehat \Omega^\dagger \Omega \tau_3) \, ,
\qquad 
&
 s {\rm Tr} (\bar c_0 \widehat \Omega^\dagger \chi \tau_3) \, .
\label{ops.3}
\end{eqnarray}
At $\hat \sigma = \hat \phi_a =0$ one gets back the operators
in eq.(\ref{ops.1}).
Notice that in order to achieve local SU(2) invariance
it is necessary to introduce interaction terms with at most three
external scalar sources. 

\begin{figure}
\begin{center}
\includegraphics[width=2.5truecm]{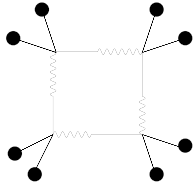}
\end{center}
\caption{Divergent one-loop graph with 
external scalar sources (denoted by lines
with a circle). 
The mixed wavy-solid lines denote $A_\mu-b$ propagators.}
\label{no.wpc}
\end{figure}

An alternative strategy is aimed at modifying the relative
coefficients of the first two and the third component of
the covariant derivative w.r.t. $V_\mu$ of $A_\mu - V_\mu$
by making use of the invariant operator
\begin{eqnarray}
s  {\rm Tr} (\bar c \widehat \Omega \tau_3 
\widehat \Omega^\dagger D_\mu[V](A-V)^\mu) \, .
\end{eqnarray}
This operator contains a smaller number of external sources, 
however it  leads to vertices with two external sources,
one Nakanishi-Lautrup field $b$ and one gauge field with one derivative.
The latter violate the WPC maximally, since they give rise
to divergent one-loop graphs with an arbitrary number of
external scalar sources of the type shown in Figure~\ref{no.wpc}.

Thus the WPC and the symmetries of the nonlinear theory
lead to the following choice
\bea
{\cal F}_a & = & D_\mu[V] (A^\mu-V^\mu)_a + \frac{f M_W}{4\xi} l_a \, , 
\label{gff.1} \\
{\cal F}_0 & = & \partial B + \frac{g'}{g}\frac{f M_W }{4 \xi} (1+ \kappa)
\frac{1}{\sqrt{1 + \frac{f^2}{v^2} C_3}} 
\Big ( \frac{f}{v} C_3 q_3 + r_3 \Big ) \, .
\label{gff.2}
\eea
In the above equation we have set
\bea
l_a = \frac{1}{\sqrt{1 + \frac{f^2}{v^2} C_b}}
\Big [ \frac{f}{v} C_b {\rm Tr}  \Big (  \widehat \Omega 
~ \tilde q_b \frac{\tau_b}{2}  ~ \widehat \Omega^\dagger  \tau_a \Big )
+  {\rm Tr}  \Big (  \widehat \Omega 
~ \tilde r_b \frac{\tau_b}{2}  ~ \widehat \Omega^\dagger  \tau_a \Big )
\Big ]
\label{la}
\eea
and
\begin{eqnarray}
\tilde q_a = e_a q_a  \, , \qquad
\tilde r_a = e_a r_a \,  ~~~~~~~ \mbox{(no summation over } a) \, .
\label{gff.2bis}
\end{eqnarray}
The coefficients $e_a$ are defined as
\begin{eqnarray}
e_a = 1 + \kappa \delta_{a3} \, ,
\label{gff.2.1}
\end{eqnarray}
with $1+\kappa =  \frac{c_W M_Z}{M_W}$.
$\xi$ is the gauge parameter. 
In terms of $A,B$ one has
$$ \kappa = \sqrt{1+\frac{f^4 B}{2 (v^2 + A f^2)}}-1 \, . $$

From eq.(\ref{la}) it is clear that 
${\cal F}_a$ transforms in the adjoint representation
of the SU(2) group, while
${\cal F}_0$ is invariant.

%

Notice that $e_3 = 1 + \kappa$ in eq.(\ref{gff.2.1}) is different
from $e_1 =e_2=1$. This is because in the
nonlinearly realized electroweak model there is an
independent mass invariant for the $Z$ vector
meson, controlled by the parameter $\kappa$.
Therefore the standard background 't Hooft gauge-fixing~\cite{SMBFM}
cannot be used here.
Moreover, it should be stressed that
the gauge-fixing functions in eqs.(\ref{gff.1}) and (\ref{gff.2}) 
are nonlinear in the quantum fields, due to the
presence of the nonlinear constraint $\phi_0$.
Nevertheless the b-equations can be written, as shown in 
eq.(\ref{b.eq}) in Appendix~\ref{app:fi}.

The gauge-fixing part is finally
\begin{eqnarray}
S_{g.f.} & = &  \int d^4x \, s \Big [ \bar c_a ( \frac{1}{4\xi} b_a + {\cal F}_a ) +
\bar c_0 ( \frac{1}{4\xi} b_0 + {\cal F}_0)  \Big ] \nonumber \\
& = & \int d^4x \, \Big [ \frac{1}{4\xi} b_a^2 
+ b_a {\cal F}_a - \bar c_a s {\cal F}_a + \frac{1}{4\xi} b_0^2 
+ b_0 {\cal F}_0 - \bar c_0 s {\cal F}_0 \Big ] \, .
\label{app.1.3}
\end{eqnarray}

At zero background fields one finds
\bea
&& 
\!\!\!\!\!\!\!\!\!\!\!\!\!\!\!\!\!\!\!\!\!\!\!\!\!
\left . {\cal F}_a  \right |_{\hat \sigma = \hat \phi_a = V_{a\mu} = 0} = 
\partial A_a + \frac{M_W}{2 \xi} \phi'_a e_a \, , 
~~~
 \left . {\cal F}_0  \right |_{\hat \sigma = \hat \phi_a= V_{a\mu} = 0} = 
\partial B + \frac{g'}{g}  \frac{M_W}{2 \xi} ( 1 + \kappa) \phi_3' \, ,
\eea
so that the gauge-fixing in eq.(\ref{app.1.3}) yields indeed
diagonal gauge and Goldstone bosons propagators in the 
't Hooft gauge. They are summarized 
in Appendix~\ref{app:thooft}.

The BRST transformations are collected in
Appendix~\ref{app:brst}.
The LFE is not spoiled provided that
$\bar c_a, b_a$ transform in the adjoint
representation of SU(2), 
while $\bar c_0, b_0, c_0$ are invariant.
Moreover the BRST partner $\Theta \equiv \frac{1}{f} ( \Theta_0 + i \Theta_a \tau_a )$ of $\widehat \Omega$
\begin{eqnarray}
s \widehat \Omega = \Theta \, , ~~~ 
s \Theta = 0 
\label{app.1.4}
\end{eqnarray}
should have the same transformation properties as $\widehat \Omega$.
The ghost equations are given in eq.(\ref{gh.eqs}).

In the BRST quantization of gauge theories~\cite{brst} the 
physical Hilbert space ${\cal H}$ is identified with
the quotient space  ${\cal H} = {\rm ker }~Q/{\rm Im} ~Q$.
$Q$ is the asymptotic BRST charge. Its action on the mass
eigenstates is obtained by keeping the linear terms
in the ghost fields of the full BRST transformation.

Since $\phi_0 = f + \dots$ and $\chi_0 = v + X_0$, one gets
\bea
&&
[ Q, \phi'_j ] = \frac{g}{2} \sqrt{v^2 + f^2 A} ~ c_j\, , ~~~ j=1,2 \, , \qquad
[ Q, \phi'_3 ] =  \frac{1}{2} \sqrt{v^2 + f^2 C_3} 
~ ( g c_3 + g' c_0) \, , \nonumber \\
&& 
[Q, \chi^{'\pm}] = 0 \, , \qquad [Q, \chi'_3] = 0 \, .
\label{ghost.1}
\eea
In the above equation $c_0$ is the ${\rm U(1)_Y}$ ghost.
Thus we see that the $\chi'_a$ belong to  ${\cal H}$
and hence describe physical scalar resonances, 
while the $\phi'_a$ are outside the physical Hilbert space.
They play the role of the unphysical Goldstone bosons.

In the SM limit $B=0$, $A\rightarrow 0$,
$\phi'_a$ reduce to $\chi_a$, as can be seen
by inverting the transformation (\ref{rot.1}).
The St\"uckelberg mass terms in
eq.(\ref{mass.nonlin}) disappear and
only the Higgs part (\ref{mass.lin}) survives.
The trace component $X_0$ of the $\chi$ doublet
is the Higgs field, while $\chi_a$ are 
the Goldstone fields.
As expected, in the limit $B=0$, $A\rightarrow 0$, which implies
$C_a \rightarrow 0$,
the asymptotic BRST symmetry in eq.(\ref{ghost.1})
reduces to the SM one on the $\phi'_a$.
From eq.(\ref{masses}) one also sees
that for $B=0$, $A\rightarrow 0$ the masses
of the scalar resonances $\chi'$ go to infinity.

\section{Weak Power-Counting}
\label{sec:wpc}

The classical action of the
nonlinearly realized electroweak theory  is gauge-invariant and respects the 
WPC condition, i.e.  only a finite number
of ancestor  amplitudes is divergent at each loop order.
An infinite number of divergent descendant 
amplitudes exists already at one loop order, however the subtraction
of the divergent ancestor amplitudes (which are in finite number
at each loop order)
is sufficient to make the theory finite recursively in the loop
expansion, since the divergences
of the descendant amplitudes are fixed by the LFE in terms of those
of the ancestor ones.

In the 't Hooft gauge one has to consider also
the sources $\hat \sigma, \hat \phi_a$ and 
their BRST partners $\Theta_0, \Theta_a$. 
The WPC condition can be derived as follows.

The superficial degree of divergence of a $n$-loop graph ${\cal G}$ can
be written as
\bea
d({\cal G})  =  n D - 2 I_{-2} - I_F + 2 V_2 + V_1
\label{wpc.1}
\eea
where $I_{-2}$
denote the number of internal lines associated with
propagators decreasing like $p^{-2}$,
$I_F$  the number of internal fermionic lines, 
$V_2$ is the number of vertices with two derivatives and
$V_1$ is the number of vertices with one derivative.
Moreover the number of internal lines is $I= I_{-2} + I_F+ I_{b'}$,
where $I_{b'}$ is the number of internal $b'$-lines.
$b'$ is the combination of $b,\partial A$ and $\phi'$ fields 
with diagonal propagators (see eq.(\ref{bprime})).

Since there are at most two derivatives in each interaction vertex,
one also has
\be
V = V_{2} + V_{1} + V_0
\ee
where $V_0$ is the number of vertices with no derivative interactions.
Euler's relation 
\be
I = n + V - 1
\ee
allows to replace  $I_{-2}$ in eq.(\ref{wpc.1}):
\bea
d({\cal G})  =  (D-2)n + 2 + I_F + 2 I_{b'} - V_1 - 2 V_0 \, .
\label{wpc.2}
\eea
Now one sees that $I_F$ fulfills the following bound:
\bea
I_F \leq V_{F \bar F \dots}
\label{wpc.3a}
\eea
where $V_{F \bar F \dots}$ denotes the number of vertices involving
a fermion, an antifermion and an arbitrary number of other legs.
Similarly 
\bea
I_{b'} \leq V_{b' \dots}
\label{wpc.3b}
\eea
Moreover, from the Feynman rules of the theory we see that the vertices $F \bar F \dots$
and $b' \dots$ do not involve derivatives, so that by using
eqs.(\ref{wpc.3a}) and (\ref{wpc.3b}) into eq.(\ref{wpc.2}) we find
\bea
d ({\cal G}) \leq (D-2) n + 2 - V_{\bar F F \dots} - V_1 - 2 V_{0[\bar F F b']} \, .
\label{wpc.bound}
\eea
In the above equation $V_{0[\bar F F b']}$ stands for the number of vertices with no derivative interactions and no $\bar F, F$ and $b'$ legs.

Among these vertices there are all those involving an antifield $\Phi^*$, 
where $\Phi^*$ runs over $A^{a*}_\mu, c^*_a, \chi^*_0, \chi^*_a, \phi^*_a, \phi^*_0$
and the fermion antifields $L^*, \bar L^*$ and $R^*, \bar R^*$. 
Clearly the number of external legs of a given antifield $\Phi^*$  
equals the number of vertices involving a $\Phi^*$ (notice that 
all interaction vertices are linear in the antifields):
\bea
N_{\Phi^*} = V_{\Phi^* \dots}
\eea
where the dots denote the quantum fields entering in the interaction vertex with the $\Phi^*$. 
A similar argument shows that the same result is true for all
external sources with linear couplings, namely  $K_0, V_\mu, \Omega_\mu$ and $
\Theta$:
\bea
N_{K_0} = V_{K_0 \dots} \, , ~~~~ N_{V_\mu} = V_{V_\mu \dots} \, , ~~~~ N_{\Omega_\mu} = V_{\Omega_\mu \dots} \, , 
~~~~ N_{\Theta} = V_{\Theta \dots} \, .
\eea
Moreover 
\bea
V_{\bar F F} \geq N_F + N_{\bar F} 
\label{wpc.ff}
\eea
where the equality holds true for those graphs where all fermion legs
are external.

The remaining vertices involving one derivative
(counted in $V_1$ in eq.(\ref{wpc.bound})) 
must be considered together with those without derivative interactions
contributing to $ V_{0[\bar F F b']}$.
The sum over the vertices is clearly greater or equal to the number 
of external ancestor legs, provided that each vertex is counted
with its multiplicity with respect to a given quantum field. The latter is defined 
as the maximum
number of external legs of a given type that can be generated
by the vertex.
For instance, the multiplicity of a quadrlinear $AAAA$ vertex
is $2$, since a vertex of this type can give rise at most 
to two external $A$-legs.

On the other hand, one can observe that the vertices
involving the external sources $\hat \phi$ contain at most three
$\hat \phi$'s external legs.
Then one gets in a straightforward way the following inequality:
\bea
&& \!\!\!\!\!\!\!\!\!\!\!\!\!\!\!\! V_1 + 2 V_{0[\bar F F b']} \geq N_A + N_B + N_\chi + N_c + N_{\bar c} + \frac{1}{3} N_{\hat{\phi}} + N_V  \nonumber \\
&& \!\!\!\!\! + N_{\phi_a^*} + 2 ( N_{\Theta} + N_{\Omega^\mu} + N_{A^*} + N_{L^*} + N_{{\bar L}^*} +
N_{R^*} + N_{{\bar R}^*} + N_{c^*} + N_{K_0} + N_{\phi_0^*}) \, .
\label{wpc.ineq}
\eea
Notice that the sources $\phi_a^*$ have a rather special couplings, since they enter into vertices
of the form $\phi_a^* c \phi^k$ and hence can yield both an external ghost leg $c$ and 
an external $\phi_a^*$-leg. This explains their different coefficient in eq.(\ref{wpc.ineq}) 
w.r.t. the other antifields.

By using eqs.~(\ref{wpc.ff}) 
and  (\ref{wpc.ineq}) into eq.(\ref{wpc.bound}) we arrive
at the WPC formula
\bea
&& \!\!\!\!\!\!\!\!\!\!\!\!\!\!\!\!\!\!
 d({\cal G})  \leq (D-2) n + 2 - N_A - N_B - N_\chi - N_F - N_{\bar F} - N_c - N_{\bar c} 
 - \frac{1}{3} N_{\hat{\phi}} - N_V  -  N_{\phi_a^*}\nonumber \\
&& - 2 ( N_{\Theta} + N_{\Omega^\mu} + N_{A^*} + N_{L^*} + N_{{\bar L}^*} +
N_{R^*} + N_{{\bar R}^*} + N_{c^*} + N_{K_0} + N_{\phi_0^*}) \, .
\label{wpc.formula}
\eea

\begin{figure}
\begin{center}
\includegraphics[width=2.5truecm]{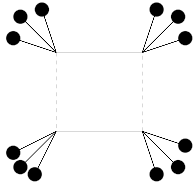}
\end{center}
\caption{Logarithmically divergent one-loop graph with 
twelve external scalar sources (denoted by lines
with a circle). Internal solid lines denote Goldstone propagators, dashed lines $b'$-propagators.}
\label{fig.0}
\end{figure}

In Fig.~\ref{fig.0} a logarithmically divergent one-loop
diagram with 12 external scalar sources is depicted.
Thus the bound in eq.(\ref{wpc.formula}) for the external scalar
sources is saturated already
at one loop.

Notice that  by eq.(\ref{wpc.formula}) the $b'$
fields have UV degree zero. The amplitudes involving the $b'$'s are
fixed by the b-equations (\ref{b.eq}).

\section{Tree-level Unitarity}\label{sec:treeunit}

In order to study the asymptotic high-energy behaviour of the nonlinearly realized electroweak theory, 
we consider the elastic scattering of longitudinally polarized gauge bosons at tree-level. 

The trilinear and quadrilinear gauge couplings of the nonlinearly realized electroweak model, which arise from the field strength term
$-\frac{1}{4}\, G_{\mu\nu}^a G^{a\mu\nu}$, coincide with the corresponding vertices of the SM. 
Thus, we need to focus only on the couplings between the scalar CP-even neutral resonance $X_0$ and the gauge bosons $W$ and $Z$. The relevant 
vertices are:
\begin{eqnarray}
\frac{g\, M_W}{\sqrt{1+A\,\frac{f^2}{v^2}}}\, X_0\, W^+ \cdot W^- + \frac{1}{2}\, \frac{G\, M_Z}{\sqrt{1+\frac{f^2}{v^2}\Big(A+\frac{B f^2}{2}\Big)}} \,
X_0\, Z^2\,.
\label{eq.ver.1}
\end{eqnarray}
We recall that in the nonlinearly realized electroweak model without physical scalar resonances \cite{Bettinelli:2008ey, Bettinelli:2008qn}
these vertices are absent, while in the SM they read
\begin{eqnarray}
g\, M_W \, X_0\, W^+ \cdot W^- + \frac{1}{2}\, G\, M_Z \,X_0\, Z^2\,.
\label{eq.ver.2}
\end{eqnarray}
Hence, in the nonlinearly realized electroweak model with scalar resonances the couplings of the gauge bosons to the CP-even resonance 
acquire a parametric dependence on $A,B$. The dependence on $A$ is 
at the origin of the  violation of the tree-level unitarity bound \cite{Froissart:1961ux, Lee:1977eg} in the elastic scattering of 
longitudinally polarized gauge bosons.

We notice that, at tree-level, the other physical scalar resonances do not contribute to the elastic scattering of gauge bosons.

\begin{figure}[p]
\begin{center}
\includegraphics[width=0.73\textwidth]{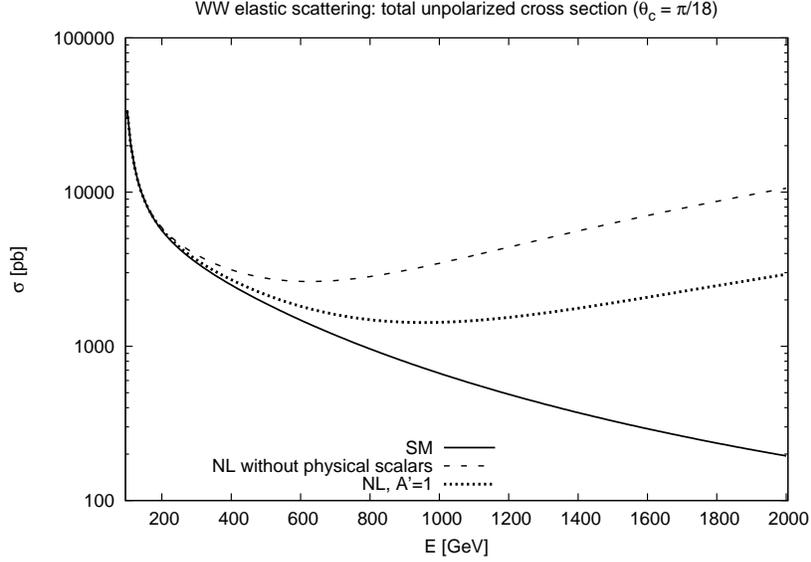}
\end{center}
\caption{Total unpolarized cross section of the elastic $WW$ scattering. The scale on the vertical axis is logarithmic.
The solid line corresponds to the SM case, the dashed one to the nonlinearly realized electroweak model without scalar resonances and finally the 
dotted line shows the result for the nonlinearly realized electroweak model with scalar resonances and $A' = 1$.}
\label{fig.1}
\end{figure}

As an example, we show in Fig.~\ref{fig.1} the total unpolarized cross section of the 
elastic scattering of charged $W$ bosons in three cases: the SM (where the mass of the gauge bosons
is generated via the Higgs mechanism), the nonlinearly realized electroweak model without scalar resonances (where the mass of the gauge bosons is
generated via the St\"uckelberg mechanism) and the nonlinearly realized electroweak model with scalar resonances where we have set $A' \equiv A\, \frac{f^2}{v^2} = 1$ 
(i.e. $50 \%$ of the mass is generated via the Higgs mechanism and $50\%$ via the St\"uckelberg one). The curves have been obtained by 
following the conventions reported in Ref.~\cite{Denner:1997kq}. In particular, in the integration over the 
two-body phase space a cut-off $\theta_c$ has been introduced, $\theta \in \big[\theta_c, \pi-\theta_c\big]$,
with $\theta_c = \frac{\pi}{18}$. One clearly sees that also the nonlinearly realized electroweak model with physical scalar
resonances exhibits a bad polynomial behaviour in the high-energy limit. 

\begin{figure}[p]
\begin{center}
\includegraphics[width=0.73\textwidth]{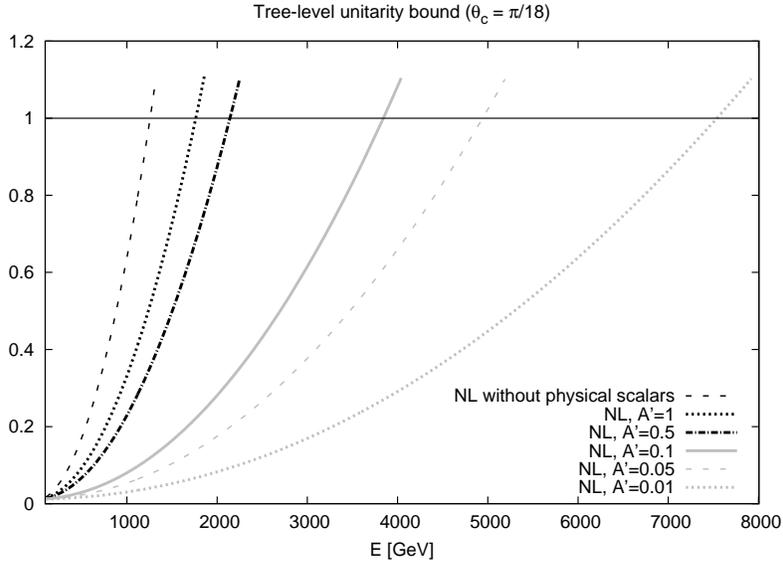}
\end{center}
\caption{Tree-level unitarity bound. The intersection between the curves and the horizontal line gives the energy threshold. The black dashed line 
shows the result for the nonlinearly realized electroweak model without scalar resonances, while the other curves correspond to the nonlinearly realized 
electroweak model with scalar resonances and various values of the $A'$ parameter, namely $A' = 1$ the black dotted one, $A' = 0.5$ the black dashed-dotted one, 
$A' = 0.1$ the grey solid one, $A' = 0.05$ the grey dashed one and finally $A' = 0.01$ the grey dotted one.}
\label{tlu}
\end{figure}

In order to assess quantitatively the energy threshold at which perturbative unitarity becomes untenable, we consider the tree-level 
unitarity bound for the elastic scattering of longitudinally polarized gauge bosons in the case of the nonlinearly realized electroweak 
model without and with physical scalar resonances.  
To this end, let $\mathcal{M}(E^2,\theta)$ be the longitudinally polarized 
scattering amplitude. We project this amplitude into partial waves
\begin{eqnarray}
\mathcal{M}^j(E^2) = \frac{1}{32 \pi}\, \int_{\theta_c}^{\pi-\theta_c}\!\! d \theta \, \sin \theta\, 
\mathcal{M}(E^2,\theta)\, P_j\big(\cos \theta\big)\,,  
\label{eq.11}
\end{eqnarray}
where $P_j\big(x\big)$ are the Legendre polynomials, with $P_0(x) = 1$, $P_1(x) = x$, and so on.
The unitarity of the scattering matrix can be translated into the following condition:
\begin{eqnarray}
\vert {\rm Re} \mathcal{M}^j(E^2)\vert \leq \frac{1}{2 \sqrt{1-\frac{M^2}{E^2}}}\,,
\label{eq.12}
\end{eqnarray}
where $M$ is the mass of the scattering particle. This condition must be valid for the complete
scattering amplitude, i.e. for the sum of all orders in perturbation theory. It is more 
useful to have a condition valid for the tree-level scattering amplitude. Conventionally, we 
assume that the tree-level scattering amplitude can be above the unitarity bound, but not too 
much, if perturbation theory is to be reliable. A sensible choice, commonly used in the literature, is
to assume that the higher perturbative orders can compensate at most $50 \%$ of the violation of the 
unitarity bound. Thus, we impose the condition:
\begin{eqnarray}
\sqrt{1-\frac{M^2}{E^2}}\, \vert {\rm Re} \mathcal{M}^{j,0}(E^2)\vert \leq 1\,.
\label{eq.13}
\end{eqnarray}
In Fig.~\ref{tlu} the l.h.s. of eq.(\ref{eq.13}) is plotted as a function of the energy and compared with $1$ in the cases of the 
nonlinearly realized electroweak model without and with scalar resonances (in the latter case some values of the parameter $A'$ have been considered).    
It turns out that the projection into $P_0(x)$ gives the most stringent bounds and thus only the corresponding curves are shown in Fig.~\ref{tlu}.  
It is quite remarkable that the energy threshold at which the violation of the bound in eq.(\ref{eq.13}) occurs is already above $2\,$ TeV when $A' = 0.5$, 
which corresponds to $2/3$ of the mass of the gauge bosons generated via the Higgs mechanism, while the threshold is slightly above $1.2\,$ TeV when 
no physical scalar resonances are added to the spectrum. By choosing a rather small value for the parameter $A'$, namely $A' = 0.01$, the violation 
of the unitarity bound is pushed at about $\simeq 7.5\,$ TeV. 

By fine-tuning the parameter $A'$, the energy threshold 
where violation of tree-level unitarity occurs
can be pushed at arbitrarily high energies.
In these regions new resonances
might show up, contributing to the unitarization of the model.

An important issue is to determine the range of allowed $A'$ values,
compatible with the current LHC data. It may happen that one can already
exclude the presence of a St\"uckelberg-generated mass fraction, thus
confirming the realization of the Higgs mechanism of electroweak SSB.
This phenomenological analysis will be presented elsewhere~\cite{phenoPaper}.

At tree-level in the nonlinearly realized electroweak model without scalar resonances there are no graphs 
contributing to the elastic scattering of $Z$ bosons. The cross section of the same process in the 
model with resonances coincides  with the SM result, modulo a
 global rescaling factor $\frac{1}{1+C'}$ where $C' \equiv \frac{f^2}{v^2}\Big(A+ \frac{B f^2}{2}\Big)$. Thus, 
the elastic scattering of longitudinally polarized  
$Z$ bosons in the nonlinearly realized electroweak model does not violate the tree-level unitarity bound.
This means that the high-energy asymptotic behaviour of the theory in the gauge boson sector is controlled by the $A$ parameter only, while 
$B$ plays no role. 

\section{Small $A$ Limit}\label{sec:smallA}

To a very good approximation one can assume that custodial symmetry
holds in the gauge boson sector and thus
set $B=0$.

Current LHC data clearly favour a scenario
where new physics contributions, 
resulting in deviations from the SM
values, are small~\cite{Giardino:2013bma}.

The physically interesting case is therefore achieved
in the small $A$ limit. It is interesting to note that
the Feynman rules, arising from the expansion
of the nonlinear constraint as a power series
in $\phi_a$'s, cannot be directly used.
In fact in the limit $A \rightarrow 0$
one gets
\bea
\phi_0 = \sqrt{f^2 - \phi_a^2} \sim f \sqrt{1 - \frac{\chi^{'2}_a}{A}} \, ,
\eea
and the power series expansion in terms of $\chi^{'2}_a / A$ cannot be
carried out.

In order to overcome this problem 
the techniques introduced in~\cite{Quadri:2006hr} prove useful.
The nonlinear constraint is introduced through a Lagrange multiplier
$B$ as follows:
\bea
S_{constraint} = \int d^4x \, B \Big [ (\sigma + f)^2 + \phi_a^2 - f^2 \Big ]
\label{constr.1}
\eea
When the equation of motion for $B$ is imposed, one recovers the 
solution of the nonlinear constraint
\bea
\phi_0 \equiv \sigma + f = \sqrt{f^2 - \phi_a^2}
\eea
The auxiliary fields $B$ and $\sigma$ are not physical. This is
most easily seen by extending the BRST symmetry through
the so-called Abelian embedding~\cite{Quadri:2006hr}.
The Abelian antighost $\bar c$ is transformed into the constraint,
while $B$ goes into the Abelian ghost $c$:
\bea
s \bar c =  (\sigma + f)^2 + \phi_a^2 - f^2  \, , ~~~~
s B = c \, , ~~~~ s c = 0 \, .
\label{embed.1}
\eea
Nilpotency is preserved since the constraint is BRST-invariant.
Then one can embed the functional $S_{constraint}$ into the
following BRST-exact functional:
\bea
S_{embed} = \int d^4x \, s (\bar c B) = \int d^4x \, 
\Big [ B (  (\sigma + f)^2 + \phi_a^2 - f^2 ) - \bar c c \Big ]
\eea
The ghost $c$ is free. Moreover at the asympotic level one gets
from eq.(\ref{embed.1}):
\bea
&& \{ Q, \bar c\} = 2 f \sigma \, , ~~~~ [Q, \sigma] =0 \, , \nonumber \\
&& [ Q, B] = c \, , ~~~~  \{Q, c\} = 0 \, .
\eea
Thus $(\bar c,\sigma)$ and $(B,c)$ are arranged in BRST doublets,
according to their unphysical nature.

The propagators generated by $S_{embed}$ are, in terms of the canonically
normalized field $\sigma = \frac{1}{\sqrt{A}} \sigma'$: 
\bea
\Delta_{BB} = -i \frac{A p^2}{4f^2}  \, ,  ~~~ \Delta_{B\sigma'} = i \frac{\sqrt{A}}{2f} \, , ~~~ \Delta_{\sigma'\sigma'} = 0 \, .
\label{prop.em}
\eea
The WPC does not hold in the sector spanned by $\sigma$. This happens
since the propagator $\Delta_{B\sigma'}$ is a constant and thus one
can construct one-loop graphs with a chain of internal $B\sigma'$-propagators and an even number of external $\sigma'$ legs that have superficial UV
degree of divergence $4$, irrespective of the number of external $\sigma'$-legs.
However, the violation of the WPC is confined to a BRST-exact sector.
Indeed, the bleached counterpart of $\sigma'$ is precisely the
nonlinear constraint
\bea
\tilde \sigma' = \frac{\sqrt{A}}{2f^2} \Big [ \sigma^2 + 2 f \sigma + \phi_a^2 \Big ] =  \frac{\sqrt{A}}{2f^2}  s \bar c \, .
\eea
Therefore $\tilde \sigma'$ and $\bar c$ form a set of coupled BRST
doublets and one is guaranteed~\cite{Quadri:2002nh} that they do not contribute
to the cohomology of the linearized ST operator. Thus 
they can appear in the counterterms only through BRST-exact terms.

In order to establish a power-counting in $A$ for the physically relevant amplitudes in the $B=0$, $A \to 0$ limit, we first notice that  in the tree-level vertex 
functional (\ref{tree.level}) there are no singular terms in the symmetric unprimed basis.

Moreover, the propagators in the primed basis are smooth for $A\rightarrow 0$,
as can be seen from eqs.(\ref{prop.em}) and (\ref{prop.2}) - (\ref{prop.4}).
As a consequence, singularities for $A \to 0$ can only arise
from the interaction terms after the field redefinition to the
mass eisgenstates basis (\ref{rot.1})
\bea
\phi_a \sim _{A \to 0} \frac{1}{\sqrt{A}}\, \chi'_a + \dots\,,
\eea
where the dots stand for terms which are subleading in the 
small A limit,
and from the replacement $\sigma =  \frac{1}{\sqrt{A}} \sigma'$
to the canonically normalized $\sigma'$ field.

In the gauge boson mass sector, $\sigma$ and $\phi_a$ enter at most quadratically in 
the St\"uckelberg mass term in eq.(\ref{mass.nonlin}) at $B=0$ 
and thus no singular interaction vertices arise.

In the physically relevant sector at zero external sources
there are no divergent terms in the gauge-fixing sector.
On the other hand, from $S_{embed}$ one gets the following singular 
interaction vertices:
\bea
\ \int d^4x \,  \Big [  \frac{1}{A} ~ B ({\sigma'}^2 + {\chi'_a}^2)
+ \frac{2}{\sqrt{A}} \frac{f}{v} B \chi'_a \phi'_a \Big ]
\eea

\begin{figure}[t]
\begin{center}
\includegraphics[width=1.0\textwidth]{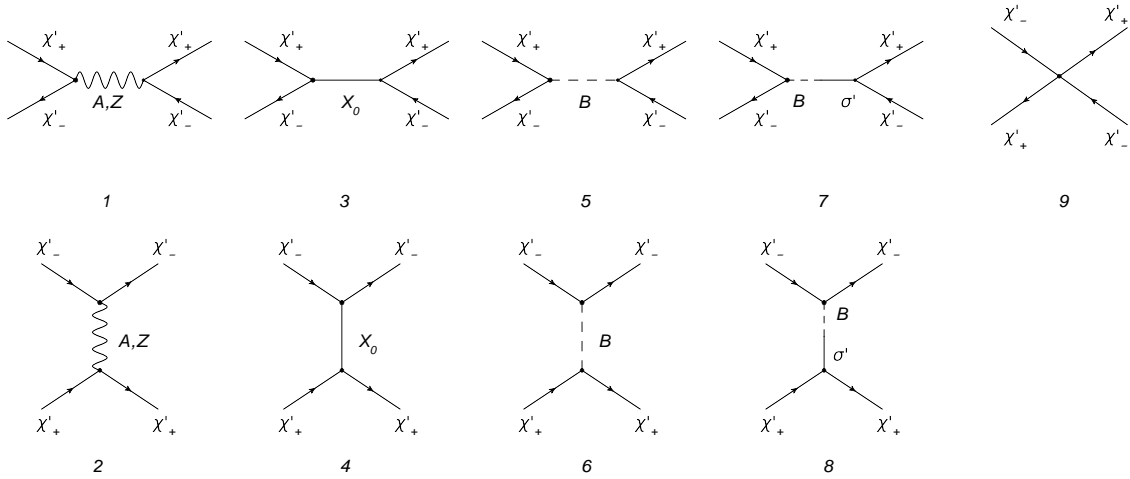}
\end{center}
\caption{Elastic scattering of charged physical scalar resonances at tree-level}
\label{fig.chi}
\end{figure}

For instance, the tree-level elastic scattering of physical scalar resonances (charged and CP-odd) is singular in the $A \to 0$ limit. 
The tree-level graphs for the elastic scattering of charged physical scalar resonances are depicted in Fig.~\ref{fig.chi}. From the previous considerations, 
it is straightforward to find the behaviour for small $A$ of the contribution of the various graphs to the scattering amplitude. In particular, the contribution coming 
from the first two graphs does not depend on $A$, the one from graphs 3 and 4 depends on $A$ through the mass of the charged scalar resonance 
($\sim M^4_{\chi'_\pm} \sim A^{-2}$), the contribution stemming from graphs 5 and 6 is singular in the small $A$ limit ($\sim A^{-1}$) and moreover it has a bad polynomial 
behaviour ($\sim s+t$) in the high energy limit, the one from graphs 7 and 8 has the same $A$ behaviour as the squared mass of $\chi'_{\pm}$ ($\sim M^2_{\chi'_\pm} \sim A^{-1}$) 
and finally the contribution coming from graph 9 vanishes when $A \to 0$ ($\sim A^2$). 
This result is consistent with the fact that the physical scalar resonances decouple in the SM limit. 

\section{Conclusions}\label{sec:concl}

A mathematically consistent nonlinearly realized
electroweak theory, incorporating physical
scalar resonances,  has been studied. It fulfills
a set of functional identities (LFE,  ST identity, b-equations,
ghost equations) as well as a
natural Hopf algebra selection criterion, embodied
in the WPC condition.
The LFE controls the deformation of the nonlinearly
realized SU(2) gauge symmetry, induced by radiative corrections, order by order in the loop
expansion. Stability of the gauge-fixing
is encoded in the b-equations and the ghost
equations. In the nonlinearly realized theory the Weinberg relation between the mass of the $Z$
and the $W$ bosons is not automatically
fulfilled and, as a consequence, two independent
mass invariants exist.
The procedure to implement the 't Hooft
gauge in the presence of two mass invariants,
while respecting the defining functional identities
and the WPC, has been given.
The ST identity in turn guarantees the fulfillment 
of physical unitarity (i.e. cancellation of 
intermediate ghost states in physical amplitudes).

The model interpolates between the Higgs and a purely
St\"uckelberg scenario. It is impossible to accommodate a single physical scalar resonance without
violating the WPC. The theory therefore
makes definite predictions on the BSM sector: 
there must be four scalar resonances, two charged
ones and two neutral ones, one CP-even (to
be eventually identified with the $125 ~GeV$
resonance recently discovered at LHC) and
one CP-odd.

We have found that if even a small fraction of the mass
is generated by the St\"uckelberg mechanism,
the Froissart bound for the scattering of
longitudinally polarized $W$ bosons is violated
at sufficiently high energies already at tree-level,
despite the exchange of a physical scalar resonance.
This feature is a characteristic footprint of
nonlinearly realized theories.

An important issue is whether one can already
exclude from the present LHC7-8 data the presence
of a St\"uckelberg component in the electroweak SSB mechanism
realized in Nature. 
As a necessary preliminary step in this direction,
we have analyzed the formal properties of the
model in the  $B=0$ (the $\rm SU(2)_R$ custodial symmetry holds in the gauge boson sector), small
$A$ limit, which is believed to be 
in a first approximation the physically
relevant scenario, since BSM effects, if present,
have to be small.
 
In this limit the Feynman rules obtained by expanding
the solution of the SU(2) nonlinear constraint in powers
of the coordinates $\phi_a$ of the SU(2) group element 
cannot be used.
In order to overcome this problem, an approach based on the
introduction of a Lagrange multiplier in the
so-called Abelian embedding formalism has been studied.
The WPC is violated, but only in an unphysical
 BRST-exact sector of the theory. 
We have also derived a power-counting to identify the leading diagrams in the small $A$ limit.
This is a necessary tool for the forthcoming
phenomenological analysis of the model.

\section*{Acknowledgments}

One of us (A.Q.) acknowledges the warm
hospitality at ECT*, Trento, where part
of this work has been carried out.
Useful discussions with D.~Binosi are gratefully acknowledged.

\appendix

\section{Conventions}

The SU(2) and ${\rm U(1)_Y}$ gauge fields are $A_{a\mu}$ and $B_\mu$ respectively.
The charged $W$, the $Z$ and photon $A$ fields
are given by
\bea
W^\pm_\mu = \frac{1}{\sqrt{2}} ( A^1_\mu \mp i A^2_\mu) \, , ~~~
Z_\mu = c_W A^3_\mu + s_W B_\mu \, , ~~~ A_\mu = - s_W A^3_\mu + c_W B_\mu \, .
\eea
In the above equation
$c_W$ and $s_W$ are the cosine and the sine of the Weinberg angle:
\bea
c_W = \frac{g}{G} \, , ~~~~ s_W = \frac{g'}{G} \, , ~~~~ G = \sqrt{g^2 + {g'}^2} \, .
\eea
$g,g'$ are the SU(2) and the hypercharge ${\rm U(1)_Y}$ coupling constants respectively.
The covariant derivative of the matrix $\chi = (v+X_0) + i \chi_a \tau_a$ is
\bea 
D_\mu \chi = \partial_\mu \chi - i g A^a_\mu \frac{\tau^a}{2} \chi  - i g' B_\mu \chi \frac{\tau_3}{2} 
\eea
and similarly for $\Omega$.

\medskip

\section{BRST Transformations}\label{app:brst}

$c_a$  are the SU(2) ghosts,
$c_0$ is the hypercharge ${\rm U(1)_Y}$ ghost.
The BRST transformations are
\bea
&& s A_{a\mu} = \partial_\mu c_a + g \epsilon_{abc} A_{b\mu} c_c \, , \qquad
s B_\mu = \partial_\mu c_0 \, ,  \qquad s c_a = -\frac{1}{2} \epsilon_{abc} c_b c_c \, ,
\nonumber \\
&& s \chi = i g c_a \frac{\tau_a}{2} \chi + i g' \chi c_0 \frac{\tau_3}{2}  , \qquad
s \Omega = i g c_a \frac{\tau_a}{2} \Omega + i g' \Omega c_0 \frac{\tau_3}{2} \, , 
\nonumber \\
&& s L = i g c_a \frac{\tau_a}{2} L + \frac{i}{2} g' c_0 Y_L L  \, , 
\qquad s R = \frac{i}{2} g' c_0 (Y_L + \tau_3) R \, , \nonumber \\
&& s \bar c_a = b_a \, , \qquad s b_a = 0 \, , \qquad s \bar c_0 = b_0 \, , \qquad
s b_0 = 0 \, \nonumber \\
&& s \phi_0^* = - K_0 \, , ~~~ s K_0 = 0 \, , ~~~
s V_{a\mu} = \Omega_{a\mu} \, , ~~~ s \Omega_{a\mu} = 0 \, ,
~~~ s \hat \Omega = \Theta \, , ~~~ s \Theta = 0 \,.
\label{brst.t}
\eea

$Y_L$ is the hypercharge of the doublet $L$.
\medskip

The BRST transformation of the background fields
$V_{a\mu}$, $\hat \Omega$ guarantee that the physical
observables of the theory are not modified~\cite{BFMtheory}, since the cohomology
of the BRST differential is unchanged by the inclusion
of the background fields as BRST doublets~\cite{Quadri:2002nh},\cite{Barnich:2000zw}.
This implies that the dependence on the background
is generated via a canonical transformation
respecting the ST identity~\cite{BFMcanonical}.

\section{Tree-level Vertex Functional}

The tree-level action of the nonlinearly realized theory is
\bea
\G^{(0)} & = & \int d^4x \, \Big (
-\frac{1}{4} G_{a\mu\nu} G_a^{\mu\nu} - \frac{1}{4} F_{\mu\nu}^2 +
\frac{1}{4} ~{\rm Tr}  (D_\mu \chi)^\dagger D^\mu \chi  \nonumber \\
& & \qquad \qquad \frac{Af^2}{4} ~{\rm Tr}(D_\mu \Omega)^\dagger D^\mu \Omega
+\frac{Bf^4}{16}~[{\rm Tr} (\Omega^\dagger D_\mu \Omega \tau_3)]^2
\nonumber \\
& & \qquad \qquad -\frac{\lambda}{16} \Big [ {\rm Tr} (\chi^\dagger \chi) \Big ]^2 +
\frac{\mu^2}{2} {\rm Tr} (\chi^\dagger \chi)  - \frac{1}{2} M_3^2 \tilde \chi_3^2 - M^2_{\pm} \tilde \chi^+ \tilde \chi^- 
\nonumber \\
& & \qquad \qquad + \sum_L \bar L ( i \slshPartial + g \slshA - \frac{g'}{2} Y_L \slshB ) L
+ \sum_R \bar R (i \slshPartial - \frac{g'}{2} (Y_L + \tau_3) \slshB) R 
\nonumber \\
& & \qquad \qquad 
+ \sum_{i,j} \Big [ \bar L^l_i Y^l_{ij} l^d_{Rj} \Xi + \bar L^q_i Y^d_{ij} q^d_{Rj} \Xi +  \bar L^q_i Y^u_{ij} q^u_{Rj} \Xi^C + h.c. \Big ] \nonumber \\
& & \qquad \qquad
+ \sum_{i,j,k} \Big [ \bar{\tilde l}^d_{Li} m^l_{ik} y^l_{kj} l^d_{Rj} +
          \bar{\tilde q}^u_{Li} m^u_{ik} y^u_{kj} q^u_{Rj} +
          \bar{\tilde q}^d_{Li} m^d_{ik} y^d_{kj} q^d_{Rj} + h.c. \Big ]
\nonumber \\
& &  \qquad \qquad  + \frac{1}{4\xi} b_a^2 
+ b_a {\cal F}_a - \bar c_a s {\cal F}_a + \frac{1}{4\xi} b_0^2 
+ b_0 {\cal F}_0 - \bar c_0 s {\cal F}_0 \nonumber \\
& &  \qquad \qquad 
+ A^*_{a\mu} s A_{a\mu} + c^*_a s c_a + \chi_0^* s \chi_0 + 
\chi_a^* s \chi_a + \phi_a^* s \phi_a + K_0 \phi_0 + \phi_0^* s \phi_0 
\nonumber \\
& & \qquad \qquad 
+ \sum_L \Big [ L^* sL + \bar L^* s\bar L \Big ] 
+ \sum_R \Big [ R^* sR + \bar R^* s \bar R \Big ]
\Big ) \, .
\label{tree.level}
\eea
In the above equation we have used the following notation for the fermions.
$L$ ranges over the set of left SU(2) doublets
\bea
L \in \Big \{ \begin{pmatrix} l^u_{Lj} \\ l^d_{Lj} \end{pmatrix} \, ,
\begin{pmatrix} q^u_{Lj} \\ q^d_{Lj} \end{pmatrix} \, , ~~~~
j=1,2,3 \Big \}
\eea
The lepton doublet of the generation $i$ is denoted by
$ L^l_i = \begin{pmatrix} l^u_{Li} \\ l^d_{Li} \end{pmatrix} $,
the quark doublet by $L^q_i = \begin{pmatrix} q^u_{Li} \\ q^d_{Li} \end{pmatrix} $.
Color indices are suppressed.
$\tilde l^{u(d)}_{Li}$ and $\tilde q^{u(d)}_{Li}$ are the components of the
bleached doublets
\bea
\tilde L^l_i = \Omega^\dagger L^l_i = \begin{pmatrix}
{\tilde l}^u_{Li} \\ {\tilde l}^d_{Li} \end{pmatrix} \, , \qquad
\qquad 
\tilde L^q_i = \Omega^\dagger L^q_i = \begin{pmatrix}
{\tilde q}^u_{Li} \\ {\tilde q}^d_{Li} \end{pmatrix} \, .
\eea
The matrices $m^{l(u,d)}$ are taken to be diagonal,
$m^{l(u,d)}_{ik} = m^{l(u,d)}_i \delta_{ik}$ (no sum over $i$).
\\

The right components are also formally arranged in doublets
\bea
R \in  \Big \{ \begin{pmatrix} l^u_{Rj} \\ l^d_{Rj} \end{pmatrix} \, ,
\begin{pmatrix} q^u_{Rj} \\ q^d_{Rj} \end{pmatrix} \, , ~~~~
j=1,2,3 \Big \}
\eea
The matrix $\chi$ is decomposed as
\bea
\chi_{\alpha\beta} = \Xi^C_\alpha \Xi_\beta \, , ~~~~
\Xi = \begin{pmatrix}i \chi_1 + \chi_2 \\ \chi_0 - i \chi_3 \end{pmatrix} \, ,
~~~
\Xi^C = i \tau_2 \Xi^* =
\begin{pmatrix} \chi_0 + i \chi_3 \\ i \chi_1 - \chi_2 \end{pmatrix} \, .
\eea

$Y^{l(u,d)}$  are external sources. They are required to have
maximal UV degree $1$. In this way the structure of the
Yukawa couplings  in eq.(\ref{tree.level}) is automatically enforced
and consequently tree-level scalar boson mediated  flavour changing neutral
currents are suppressed~\cite{Binosi:2012cz}.
Yukawa interactions and fermion masses are induced by the 
shift $Y^{l(u,d)} \rightarrow y^{l(u,d)} + Y^{l(u,d)}$.
After diagonalization of $y^{l(u,d)}$, the masses of the fermions become
\bea
m_i^{l(u,d)'} = ( m^{l(u,d)}_i + v ) y^{l(u,d)}_i \, .
\eea
The term proportional to $v$ is the same as in the SM.
The parameters $ m^{l(u,d)}_i$ are not fixed by symmetry
requirements or the WPC. They arise since the gauge symmetry
is nonlinearly realized and therefore the bleaching procedure
can be used to add independent fermion mass invariants. 
$m^{l(u,d)}_i$ are expected to be small: in the limit $B=0$, small $A$, 
one may take 
$m^{l(u,d)}_i \sim \sqrt{A} v$. 
With this choice no divergent vertices arise  in the fermionic sector
in the limit
$A \rightarrow 0$.

\section{'t Hooft Gauge-fixing in the Nonlinearly Realized Theory}\label{app:thooft}

We summarize here the propagators in the 't Hooft gauge.
The gauge-fixing functions in eqs.(\ref{gff.1})
and (\ref{gff.2}) read
in components
\begin{eqnarray}
{\cal F}_a &=& 
\partial A_a - \partial V_a + g \epsilon_{abc}
V_{b\mu} A^\mu_c \nonumber \\
&& + \frac{M_W}{4 \xi f}
 \frac{1}{\sqrt{1 + \frac{f^2}{v^2}\, C_b}} 
\Big [
\delta_{ab} ( \hat \phi_0^2 - \hat \phi_c^2) \Big (  \frac{f}{v} C_b  \tilde q_b + \tilde r_b \Big )  +
2 \hat \phi_a \hat \phi_b 
\Big (  \frac{f}{v} C_b  \tilde q_b + \tilde r_b \Big )
 \nonumber \\
&& \qquad \qquad \qquad  + 2
\epsilon_{abc} \hat \phi_0 \Big (  \frac{f}{v} C_b  \tilde q_b + \tilde r_b \Big ) \hat \phi_c  \Big ] \, , 
\nonumber \\
{\cal F}_0 & = & \partial B +
\frac{g'}{g} \frac{M_W}{2 f \xi} (1 + \kappa)
\frac{1}{\sqrt{1+ \frac{f^2}{v^2} C_3}}
\Big [ \hat \phi_0 
\Big ( \frac{f}{v} C_3 \phi_3 + \chi_3 \Big ) 
 - \hat \phi_3 \Big ( \frac{f}{v} C_3 \phi_0 + \chi_0 \Big ) \nonumber \\
&& \qquad \qquad \qquad
+\epsilon_{3bc} \hat \phi_b 
\Big (\frac{f}{v} C_3 \phi_c + \chi_c \Big ) 
 \Big ] \, .
\label{prop.001}
\end{eqnarray}

To diagonalize the quadratic part of the tree-level vertex functional
at zero background fields
we perform the following transformation:
\bea
b'_a = b_a + 2\xi \Big (\partial A_a + \frac{M_W}{2\xi} e_a \phi'_a \Big ) \, , 
~~~~~
b'_0 = b_0 + 2 \xi 
\Big [ \partial B + \frac{g'}{g} \frac{M_W}{2\xi}(1+\kappa) \phi'_3 \Big ] \, .
\label{bprime}
\eea
We give here the propagators in the symmetric basis,
which is most useful in establishing the WPC:
\begin{eqnarray}
&& \Delta_{A_{1\mu} A_{1\nu}} = \Delta_{A_{2\mu} A_{2\nu}} = 
\frac{i}{-p^2 + M_W^2} T_{\mu\nu} 
+  \frac{i}{-2 \xi p^2 + M_W^2} L_{\mu\nu}
\, , \nonumber \\
&&
\Delta_{A_{1\mu}A_{2\nu}}=  \Delta_{A_{1\mu} A_{3\nu}} = \Delta_{A_{2\mu} A_{3\nu}} = 0 \, , 
\nonumber \\
&& \Delta_{A_{3\mu} A_{3\nu}} =  i
T_{\mu\nu} \Big ( \frac{s_W^2}{-p^2}
+ \frac{c_W^2}{-p^2 + M_Z^2} \Big )
+ i L_{\mu\nu} \Big ( 
\frac{s_W^2}{2 \xi p^2} + 
\frac{c_W^2}{-2 \xi p^2 + M_Z^2} \Big ) \, ,
\nonumber \\
&&
\Delta_{A_{3\mu} B_\nu} = -
i c_Ws_W T_{\mu\nu}
\Big ( -\frac{1}{p^2} - \frac{1}{-p^2 + M_Z^2} \Big )
- i c_Ws_W L_{\mu\nu}
\Big ( \frac{1}{2 \xi p^2} - 
\frac{1}{-2 \xi p^2 + M_Z^2} \Big )
\, , \nonumber \\
 && \Delta_{B_\mu B_\nu} = 
i T_{\mu\nu} \Big ( \frac{c_W^2}{-p^2}
+ \frac{s_W^2}{-p^2 + M_Z^2} \Big )
 + i L_{\mu\nu} \Big ( 
\frac{c_W^2}{2 \xi p^2} + 
\frac{s_W^2}{-2 \xi p^2 + M_Z^2} \Big ) \, ,
\nonumber \\
\nonumber \\ 
&& \Delta_{A_{1\mu} B_\nu} = \Delta_{A_{2\mu} B_\nu} = 0 \, ,
~~~~~
\Delta_{\phi'_i \phi'_j} = \delta_{ij} \frac{i}{p^2 
- \frac{M_W^2}{2\xi}} \, , ~~~ i,j=1,2 \, , ~~~~
\Delta_{\phi'_3 \phi'_3} = \frac{i}{p^2 - \frac{M_Z^2}{2\xi}} \, ,
\nonumber \\
&&
\Delta_{b'_a b'_b} = 2 i \xi \delta_{ab} \, , ~~~~ 
a,b=1,2,3 \, , ~~~~ 
\Delta_{b'_0 b'_0} = 2 i \xi \, , \nonumber \\
&&
\Delta_{\bar c_i c_j}  = \delta_{ij} \frac{i}{p^2 -
\frac{M_W^2}{2\xi} } \, , ~~~~ i,j=1,2 \, , 
~~~~
\Delta_{\bar c_3 c_3} = \frac{i}{p^2 - \frac{M_Z^2}{2\xi}} \, , ~~~ 
\Delta_{\bar c_0 c_0} = \frac{i}{p^2} \, . 
\label{prop.2}
\end{eqnarray}
The mixed $A-\phi$ propagators are zero. Fermion propagators
are the usual ones
\be
\Delta_{\bar f f}= \frac{i}{\slshp - m'_f}
\label{prop.3}
\ee
while for the $\chi'_a$
%

\be
\Delta_{\chi'_i \chi'_j} = \delta_{ij} \frac{i}{p^2 - M^2_{\chi'_\pm}} \, , ~~ i,j=1,2, ~~~
\Delta_{\chi'_3 \chi'_3} =  \frac{i}{p^2 - M^2_{\chi'_3}} \, , ~~~
\Delta_{X_0 X_0} = \frac{i}{p^2 - M^2_{X_0}} \, .
\label{prop.4}
\ee

\section{Functional Identities}\label{app:fi}

We collect here in compact form the functional identities of the theory:
\begin{itemize}
\item the b-equations
\bea
&& \frac{\delta \G}{\delta b_a} = \frac{1}{2\xi} b_a + \left . 
{\cal F}_a \right |_{\phi_0 \rightarrow \frac{\delta \G}{\delta K_0}} \, , 
\nonumber \\
&&  \frac{\delta \G}{\delta b_0} = \frac{1}{2\xi} b_0 + \left . 
{\cal F}_0 \right |_{\phi_0 \rightarrow \frac{\delta \G}{\delta K_0}} \, .
\label{b.eq}
\eea
Notice that the r.h.s. of the above equations is linear in the quantum fields.
The nonlinear constraint $\phi_0$ is generated by taking a derivative
w.r.t. the external source $K_0$.
\item the ghost equations 
\bea
\frac{\delta \G}{\delta {\bar c}_a} & = &
-D^\mu[V]_{ab} \frac{\delta \G}{\delta A^{*b}_\mu} 
+ D_\mu[A]_{ab} \Omega^\mu_b \nonumber \\
&& - \frac{M_W}{4 \xi f}\frac{1}{\sqrt{1 + \frac{f^2}{v^2} C_b}}
\Big \{
\frac{f}{v} C_b {\rm Tr} \Big [ \hat \Omega \Big ( \hat \phi_0 \frac{\delta \G}{\delta \phi_b^*} - \hat \phi_b \frac{\delta \G}{\delta \phi_0^*} 
-\epsilon_{bdc} \frac{\delta \G}{\delta \phi^*_d} \hat \phi_c \Big ) e_b \tau_b \hat \Omega^\dagger \tau_a \Big ] 
\nonumber \\
&& \qquad \qquad \qquad \quad ~ ~~~~~
+ {\rm Tr} \Big [ \hat \Omega \Big ( \hat \phi_0 \frac{\delta \G}{\delta \chi_b^*} - \hat \phi_b \frac{\delta \G}{\delta \chi_0^*} 
-\epsilon_{bdc} \frac{\delta \G}{\delta \chi^*_d} \hat \phi_c \Big ) e_b \tau_b \hat \Omega^\dagger \tau_a \Big ] 
\nonumber \\
&&  \qquad \qquad \qquad \quad ~ + \frac{f}{v} C_b {\rm Tr} \Big [ 
 \hat \Omega \Big (
\Theta_0 \phi_b - \Theta_b \frac{\delta \G}{\delta K_0} - \epsilon_{bdc} \phi_d \Theta_c \Big ) e_b \tau_b \hat \Omega^\dagger \tau_a \Big ]
\nonumber \\
&&  \qquad \qquad \qquad \quad ~~~~~ ~ 
+ {\rm Tr} \Big [  \hat \Omega \Big ( \Theta_0 \chi_b - \Theta_b \chi_0 - \epsilon_{bdc} \chi_d \Theta_c \Big ) e_b \tau_b \hat \Omega^\dagger \tau_a \Big ] \Big \} \nonumber \\
&&
- \frac{f M_W}{4 \xi \sqrt{1 + \frac{f^2}{v^2} C_b}}
\Big \{ 
\frac{f}{v} C_b {\rm Tr} 
\Big [
  \Theta
~ \tilde q_b \frac{\tau_b}{2}  ~ \widehat \Omega^\dagger  \tau_a 
+
\widehat \Omega 
~ \tilde q_b \frac{\tau_b}{2}  ~ \Theta^\dagger  \tau_a 
\Big ]
\nonumber \\
&&  \qquad \qquad \qquad ~~ ~~~~~ +  {\rm Tr}  \Big [  \Theta 
~ \tilde r_b \frac{\tau_b}{2}  ~ \widehat \Omega^\dagger  \tau_a 
+ \widehat \Omega 
~ \tilde r_b \frac{\tau_b}{2}  ~ \Theta^\dagger  \tau_a 
\Big ]
\Big \} \, ,
\nonumber \\ 
\frac{\delta \G}{\delta {\bar c}_0} & = &
- \square c_0 - \frac{g'}{g} \frac{M_W}{2 f \xi} 
\frac{1 + \kappa}{\sqrt{1+\frac{f^2}{v^2} C_3}} \times \nonumber \\
&&  \qquad
\Big [ \frac{f}{v} C_3 \Big (\hat \phi_0 \frac{\delta \G}{\delta \phi_3^*} - \hat \phi_3 \frac{\delta \G}{\delta \phi_0^*} 
-\epsilon_{3dc} \frac{\delta \G}{\delta \phi^*_d} \hat \phi_c \Big )
+ \frac{f}{v} C_3 \Big ( \Theta_0 \phi_3 - \Theta_3 \frac{\delta \G}{\delta K_0}
 -\epsilon_{3dc} \phi_d \Theta_c \Big )
 \nonumber \\
&&  \qquad ~~~ + \hat \phi_0 \frac{\delta \G}{\delta \chi_3^*} - \hat \phi_3 \frac{\delta \G}{\delta \chi_0^*} 
-\epsilon_{3dc} \frac{\delta \G}{\delta \chi^*_d} \hat \phi_c 
+ \Theta_0 \chi_3 - \Theta_3 \chi_0 - \epsilon_{3dc} \chi_d \Theta_c
\Big ] \, .
\label{gh.eqs}
\eea
\item the Local Functional Equation
\begin{eqnarray}
\!\!\!\!\!\!\!\!\!
({\cal W}\G)_{a} & \equiv & 
-\partial_\mu \frac{\delta \G}{\delta  V_{a \mu}} 
+ g\epsilon_{abc} V_{c\mu} \frac{\delta \G}{\delta V_{b\mu}}
-\partial_\mu \frac{\delta \G}{\delta A_{a \mu}} 
+ g\epsilon_{abc} A_{c\mu} \frac{\delta \G}{\delta A_{b\mu}}
\nonumber \\&& 
+ \frac{g}{2} K_0\phi_a
+ \frac{g}{2} \frac{\delta \G}{\delta K_0} 
\frac{\delta \G}{\delta \phi_a} +  
\frac{g}{2} \epsilon_{abc} \phi_c \frac{\delta \G}{\delta \phi_b} 
+ \frac{g}{2} \chi_0
\frac{\delta \G}{\delta \chi_a} +  
\frac{g}{2} \epsilon_{abc} \chi_c \frac{\delta \G}{\delta \chi_b}
\nonumber \\
&&  
+ g \epsilon_{abc} b_c \frac{\delta \G}{\delta b_b}
+ g \epsilon_{abc} \bar c_c \frac{\delta \G}{\delta  \bar c_b}
      + g \epsilon_{abc} c_c \frac{\delta \G}{\delta c_b}
 \nonumber \\&&
{
+\frac{i}{2} g\tau_aL\frac{\delta \G}{\delta  L}
-\frac{i}{2} g\bar L\tau_a\frac{\delta \G}{\delta \bar  L}
-\frac{i}{2} gL^*\tau_a\frac{\delta \G}{\delta  L^*}
+\frac{i}{2} g \tau_a\bar L^*\frac{\delta \G}{\delta \bar  L^*}
}
 \nonumber \\
&& + g \epsilon_{abc} \Theta_{c\mu} \frac{\delta \G}{\delta \Theta_{b\mu} }
      + g \epsilon_{abc} A^*_{c\mu} \frac{\delta \G}{\delta A^*_{b\mu}}
      + g \epsilon_{abc} c^*_c \frac{\delta \G}{\delta  c^*_b} 
\nonumber \\
&&    - \frac{g }{2} \phi_0^* \frac{\delta \G}{\delta \phi^*_a}+
\frac{g}{2} \epsilon_{abc} \phi^*_c \frac{\delta \G}{\delta \phi^*_b} 
+ \frac{g}{2} \phi_a^* \frac{\delta \G}{\delta \phi_0^*}
 - \frac{g }{2} \chi_0^* \frac{\delta \G}{\delta \chi^*_a}+
\frac{g}{2} \epsilon_{abc} \chi^*_c \frac{\delta \G}{\delta \chi^*_b} 
+ \frac{g}{2} \chi_a^* \frac{\delta \G}{\delta \chi_0^*}
\nonumber \\
&&
 - \frac{g }{2} \Theta_0 \frac{\delta \G}{\delta \Theta_a}+
\frac{g}{2} \epsilon_{abc} \Theta_c \frac{\delta \G}{\delta \Theta_b} 
+ \frac{g}{2} \Theta_a \frac{\delta \G}{\delta \Theta_0}
= 0 \, .
\label{bkgwi}
\end{eqnarray}
\item the Slavnov-Taylor identity
\bea
{\cal S}(\G) & = & 
\int d^4x \, \Big ( \frac{\delta \G}{\delta A^*_{a\mu}} 
\frac{\delta \G}{\delta A^a_\mu} +  \frac{\delta \G}{\delta \chi^*_a} \frac{\delta \G}{\delta \chi_a}
+ \frac{\delta \G}{\delta \chi^*_0} \frac{\delta \G}{\delta \chi_0}
+ \frac{\delta \G}{\delta F^*} \frac{\delta \G}{\delta F}
+ \frac{\delta \G}{\delta \bar F^*} \frac{\delta \G}{\delta \bar F}
\nonumber \\
&& \qquad \qquad + \partial_\mu c_0 \frac{\delta \G}{\delta B_\mu} 
+ \frac{\delta \G}{\delta c_a^*} \frac{\delta \G}{\delta c_a}
+ \frac{\delta \G}{\delta \phi_a^*} \frac{\delta\G}{\delta \phi_a}
+ b_a \frac{\delta \G}{\delta \bar c_a} + b_0 \frac{\delta \G}{\delta \bar c_0}
\nonumber \\
&& \qquad \qquad - K_0 \frac{\delta \G}{\delta \phi_0^*} 
+ \Omega_{a\mu} \frac{\delta \G}{\delta V_{a\mu}} 
+ \Theta_0 \frac{\delta \G}{\delta \hat \phi_0} + 
\Theta_a \frac{\delta \G}{\delta \hat \phi_a}
\Big ) = 0
\label{sti}
\eea
where $F$ runs over $L,R$.
\end{itemize}

\end{document}